\documentclass{vgtc}                          

\graphicspath{{figures/}{pictures/}{images/}{./}} 

\usepackage{times}                     

\usepackage{tabu}                      
\usepackage{booktabs}                  
\usepackage{lipsum}                    
\usepackage{mwe}                       
\usepackage{enumitem}
\usepackage{comment}
\usepackage{bm}

\usepackage{amsmath}
\DeclareMathOperator*{\ML}{ML}

\usepackage{mathptmx}                  
\usepackage[normalem]{ulem}

\onlineid{0}

\vgtccategory{IEEE VIS Short Paper}

\vgtcinsertpkg

\newenvironment{myitemize}{
\begin{itemize}
 \setlength{\itemsep}{1pt}
 \setlength{\parskip}{0pt}
 \setlength{\parsep}{0pt}}{\end{itemize} 
}

\title{FAVis: Visual Analytics of Factor Analysis for Psychological Research}

\author{Yikai Lu\thanks{e-mail: ylu22@nd.edu} %
\and Chaoli Wang\thanks{e-mail: chaoli.wang@nd.edu}} 
\affiliation{\scriptsize University of Notre Dame}

\teaser{
  \vspace{-0.1in}
  \centering
  \includegraphics[width=1.0\linewidth]{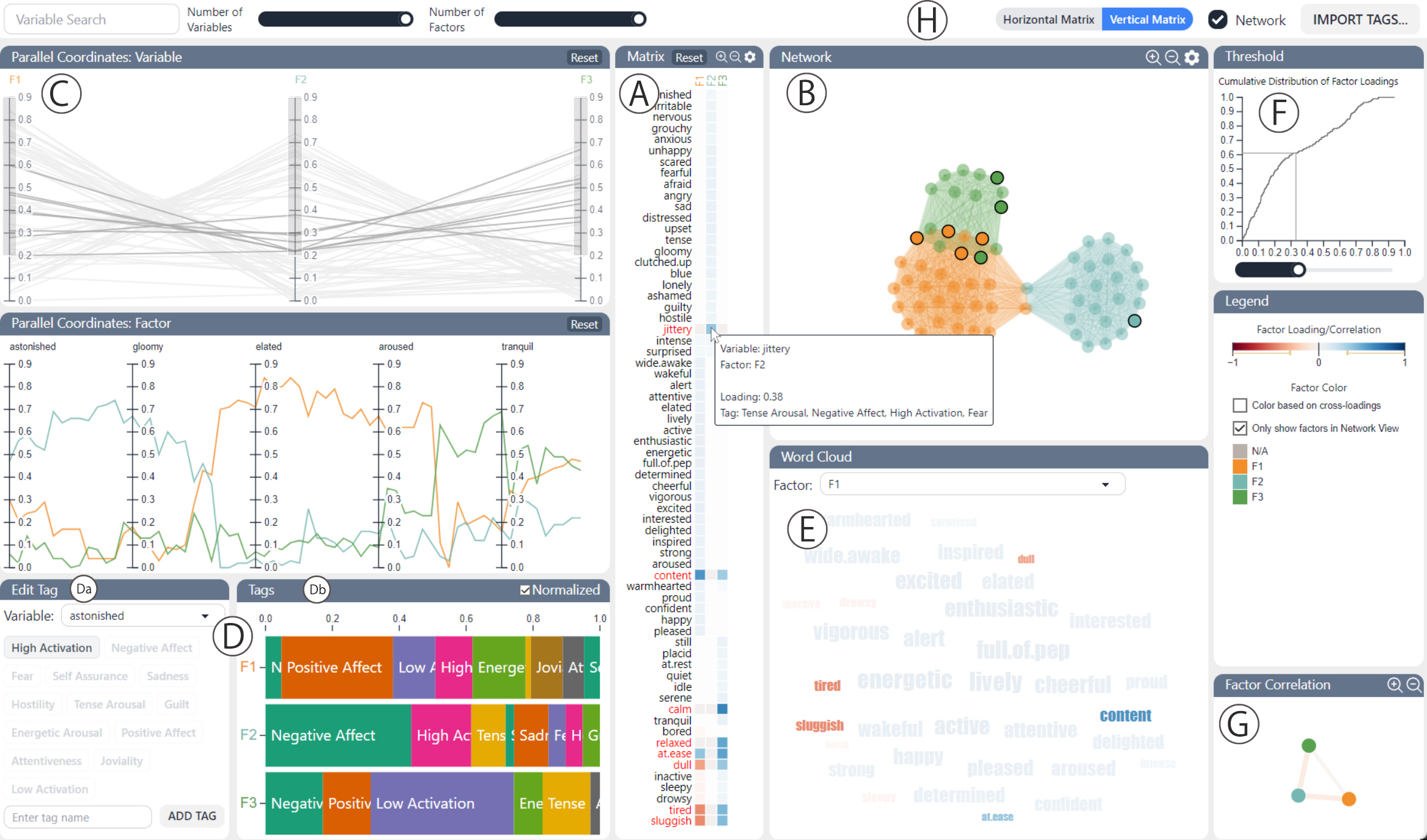}
  \vspace{-0.25in}
  \caption{FAVis (\url{https://luyikei.github.io/favis/}) visualizes a three-factor common factor model of the Motivational State Questionnaire dataset, selecting variables with the absolute values of their factor loadings larger than 0.3. (A) Matrix view shows the most direct visualization of a factor loadings matrix; (B) Network view visualizes cross-loadings most effectively and reduces redundant information; (C) Parallel-coordinates view shows factor loadings for each variable/factor without information loss and allows for selecting variables/factors in a range of values; (D) Tag view shows the summary of the relevance of tags for each factor by counting the frequency of tags annotated for variables based on a theory; (E) Word cloud view helps interpret factors by correlating fonts with the values of factor loadings; (F) Threshold view controls the number of factor loadings shown in the matrix and network views; (G) Factor correlation view shows the network of factor correlations; (H) Top bar allows users to filter variables by their prefix and limit the number of variables/factors shown in the application.}
  \label{fig:teaser}
}

\abstract{Psychological research often involves understanding psychological constructs through conducting factor analysis on data collected by a questionnaire, which can comprise hundreds of questions. Without interactive systems for interpreting factor models, researchers are frequently exposed to subjectivity, potentially leading to misinterpretations or overlooked crucial information. This paper introduces FAVis, a novel interactive visualization tool designed to aid researchers in interpreting and evaluating factor analysis results. FAVis enhances the understanding of relationships between variables and factors by supporting multiple views for visualizing factor loadings and correlations, allowing users to analyze information from various perspectives. The primary feature of FAVis is to enable users to set optimal thresholds for factor loadings to balance clarity and information retention. FAVis also allows users to assign tags to variables, enhancing the understanding of factors by linking them to their associated psychological constructs. 
Our user study demonstrates the utility of FAVis in various tasks.} 

\begin{document}



\vspace{-0.075in}
\firstsection{Introduction}

\maketitle
Psychology studies human characteristics and behavior. Unlike disciplines such as physics, where phenomena can be directly measured and observed, psychology often faces challenges in directly measuring aspects like personality, emotion, or intelligence \cite{larsenPromisesProblemsCircumplex1992,sotoNextBigFive2017, Thayer,watsonDevelopmentValidationBrief1988,willisFactorAnalyticModelsIntelligence2011}. These qualities are termed {\em psychological constructs}~\cite{crockerIntroductionClassicalModern2008a}. The primary goal of research in psychology is often to study these constructs. Although psychological constructs are unobservable, they can be inferred from observed variables. In statistical terms, psychological constructs are construed as {\em latent variables}~\cite{cameron_foundations_2010}. Latent variable models, including the {\em common factor model} (CFM)~\cite{cameron_foundations_2010,tippingProbabilisticPrincipalComponent1999a}, are employed for statistical inference.

Mathematically, let $\mathbf{x} = (x_1, \ldots, x_p)$ be a column vector representing $p$ observed variables, and $\mathbf{y} = (y_1, \ldots, y_q)$ be a column vector denoting $q$ common factors (or latent variables in broader terms), where typically $q \ll p$. The CFM is described as follows
\begin{equation}
	\mathbf{y} \sim N_q(0, \bm{\Phi}),
 \label{eqn:y}
\end{equation}
where $\bm{\Phi}$ is the $q \times q$ factor correlation matrix. Additionally,
\begin{equation}
	\mathbf{x} = \bm{\Lambda} \mathbf{y} + \bm{\mu} + \bm{\epsilon},
 \label{eqn:x}
\end{equation}
where $\bm{\Lambda}$ is a $p \times q$ matrix of factor loadings, $\bm{\mu}$ accounts for a non-zero mean of $\mathbf{x}$, and $\bm{\epsilon}$, the noise of the model, follows $\bm{\epsilon} \sim N_p(0, \bm{\Psi})$, with $\bm{\Psi}$ being a $p \times p$ diagonal matrix representing unique variances. 
The CFM is viewed as a method for dimensionality reduction, with $\mathbf{y}$ explaining most of the covariance in $\mathbf{x}$. From Equations~\ref{eqn:y}~and~\ref{eqn:x}, it follows that
\begin{equation}
	\mathbf{x} \sim N_p(\bm{\mu}, \bm{\Lambda}\bm{\Phi}\bm{\Lambda}^{T} + \bm{\Psi}).
\end{equation}
To enhance the interpretability of the parameters, including $\bm{\Lambda}$, $\bm{\Phi}$, and $\bm{\Psi}$, they can be refined through factor rotation by putting some sparsity criteria on $\bm{\Lambda}$ without altering the model.

{\em Factor analysis} (FA) is the method used to estimate the CFM. This paper focuses on {\em exploratory factor analysis} (EFA)~\cite{cameron_foundations_2010}, a technique where no predefined structure is imposed on the factor loadings. There are variants of FA or CFM. For instance, probabilistic principal component analysis (PCA)~\cite{tippingProbabilisticPrincipalComponent1999a} assumes equal unique variances ($\psi_i = \sigma^2$). In this scenario, the maximum likelihood estimators $\Lambda_{\ML}$ and $\sigma^2_{\ML}$ for the isotropic error model correspond to PCA. Consequently, our visualization tool can also assist in interpreting PCA results.

Psychologists utilize the CFM to identify psychological constructs (i.e., estimating $\bm{\Lambda}$ from data). Typically, researchers create observed variables like test questions in a relevant domain and apply FA to the data. They often standardize $\mathbf{x}$ to ensure uniform scaling. Then, they examine the factor loadings matrix $\bm{\Lambda}$ to determine which dimensions of $\mathbf{y}$ correspond to the psychological constructs.

For instance, a study measured two soldiers' psychological constructs: fear response and optimism~\cite{cameron_foundations_2010}. The researchers developed 14 seven-point adjectival rating scales for measurement. After data collection, standardization, and FA, they obtained an initial factor loadings matrix $\bm{\Lambda}_1$ with four common factors and an identity matrix for $\Phi$. However, after they attempted to interpret $\bm{\Lambda}_1$ by reading its values, the last two factors in $\bm{\Lambda}_1$ were labeled as ``difficult to interpret,'' leading them to apply varimax rotation to the first two factors, resulting in $\bm{\Lambda}_2$.
%
This process can be subjective and challenging, especially with a large $p$. The example illustrates the existence of ``difficult to interpret'' factors, where simply reading numbers for interpretation is demanding.

Evaluating the quality of a CFM based on its results is crucial. In psychology, an ideal factor loading structure is known as a ``simple structure''~\cite{cameron_foundations_2010}, characterized by minimal cross-loadings. {\em Cross-loadings} refer to instances where a variable has multiple large (more than two) factor loadings \cite{costelloBestPracticesExploratory2019,Tabachnick_Fidell_2007}. Suppose $\lambda_{ik}$ is the factor loading for variable $i$ and factor $k$. $\lambda_{ik}$ and $\lambda_{il}$ are considered as cross-loadings if both absolute values are large. Additionally, this paper introduces a concept deemed more problematic than cross-loadings: redundant-loadings. {\em Redundant-loadings} occur when multiple variables have large loadings on multiple factors. Specifically, $\lambda_{ik}$, $\lambda_{il}$, $\lambda_{jk}$, and $\lambda_{jl}$ are considered redundant-loadings if all of them are large. This makes interpreting factors more difficult, as variables $i$ and $j$ do not contribute to interpreting factors $k$ and $l$ due to a lack of differentiation. Then, how do we decide if a factor loading is \textit{large}? Traditionally, a threshold is used to interpret factor loadings, and various standards or methods are proposed to choose an optimal value~\cite{Tabachnick_Fidell_2007,Hair_C,maccallumSampleSizeFactor1999}. The subjective nature of this process influences the interpretation of the CFM.

In this paper, we design FAVis, a visualization tool to aid in interpreting and evaluating FA results. FAVis features multiple interactive views, including matrix, network, and parallel coordinates, to help users understand different aspects of the model parameters. Our tool could reveal insights from ``difficult to interpret'' factors, avoiding omitting important clues in the model and data. This tool was implemented using D3~\cite{d3} and Vue.js~\cite{Vue}.


\vspace{-0.075in}
\section{Related Works}
\label{sec:rw}




\textbf{Visualizing high-dimensional data.} 
Our work intersects with the visualization of high-dimensional data, mainly because FA is a dimensionality reduction method that extracts latent variables that explain observed variables. This aligns with numerous studies in visual analytics research focusing on high-dimensional data~\cite{liuVisualizingHighDimensionalData2017, sachaVisualInteractionDimensionality2017}. Various visualization tools and methods, especially those utilizing PCA, have been developed, such as iPCA~\cite{jeongIPCAInteractiveSystem2009} and BaVA~\cite{houseBayesianVisualAnalytics2015}. These approaches, however, differ in their treatment of PCA and dimensionality reduction. For instance, iPCA employs PCA primarily as a data preprocessing technique, with its main focus being the visualization of the original data points in a projection view. This contrasts with our FAVis, where the emphasis is on interpreting the model and factor loadings themselves, treated as input data. While some methods aim to improve model parameters through visual analytics~\cite{houseBayesianVisualAnalytics2015, endertObservationlevelInteractionStatistical2011}, they do not typically focus on the visualization of these parameters for interpretative purposes. FAVis, therefore, occupies a unique space in dimensionality reduction by concentrating on the interpretation of model parameters.



\textbf{Visualizing psychological data.}  
Visualization in psychological data, particularly in cognitive psychology and neuroscience, is well-developed. Existing research primarily focuses on EEG~\cite{jiVisualExplorationDynamic2019, slaybackNovelMethodsEEG2018} and fMRI data~\cite{rasheedSubjectSpecificBrainActivity2022, deridderReviewOutlookVisual2018}. For instance, Ji et al.\ \cite{jiVisualExplorationDynamic2019} developed an interactive visualization for dynamic EEG coherence networks. Similarly, a combination of visual analytics and scientific visualization has been applied in fMRI research~\cite{deridderTemporaltracksVisualAnalytics2017, deridderExplorationVirtualAugmented2015}.

However, psychological data in other domains, such as personality, educational, and developmental psychology, often rely on behavioral measures like tests, psychological scales, and surveys. FA is a standard tool for analyzing such data. One of these few examples is qgraph~\cite{qgraph}, which is an R package to visualize the CFM using graphs. However, qgraph only provides R functions to generate such visualization, thus lacking interactivity. Our FAVis aims to bridge this gap in the visual analytics literature, providing an interactive tool for these other psychology domains.


\vspace{-0.075in}
\section{Design Goals}

We discussed with two experts in quantitative psychology and conducted a literature review (including Section~\ref{sec:rw} and interactive visual analytics tools \cite{bayrakPRAGMAInteractivelyConstructing2020, munechikaVisualAuditorInteractive2022}) to derive the following design goals.
\begin{myitemize}
\vspace{-0.05in}
	\item \textbf{G1 -- Facilitate understanding of the associations between variables and factors.} The factor loadings matrix $\bm{\Lambda}$ describes how much a factor is related to a variable (as in Equation~\ref{eqn:x}), which can be hundreds or thousands of numbers depending on the number of variables and factors. We designed several means to visualize the matrix to reduce subjectivity in interpreting the factor loadings matrix by only highlighting relevant elements according to the threshold. 
	\item \textbf{G2 -- Provide useful between-factor information that affects the quality of a CFM.} As it is desirable that a CFM has a simple structure, the tool should support highlighting important cross-loadings. Also, an oblique rotation could cause factors to be correlated (i.e., the factor correlation matrix $\bm{\Phi}$ is not identical). Thus, the design should ensure that any crucial between-factor association will not be overlooked.
	\item \textbf{G3 -- Identify a threshold for factor loadings that balances interpretability and information loss.} Traditionally, a threshold is often used to facilitate the interpretation of the factor loadings matrix. With visualization, the design should lead users to determine an optimal threshold specific to a mode or dataset.
	\item \textbf{G4 -- Filter and sort variables and factors based on desirable characteristics.}  When interpreting factors, a researcher might want to know which variables have large loadings on a specific factor and vice versa.
	\item \textbf{G5 -- Provide a mechanism to relate a theory to a CFM.} Researchers often have a theory of how variables are related to psychological constructs before they conduct a study to collect data and construct a model. The design should be able to connect variables to presumed psychological constructs and effectively visualize the degree to which factors represent the psychological constructs to reduce the subjectivity of interpreting factors in terms of a theory. 
\vspace{-0.05in} 
\end{myitemize}

\vspace{-0.075in} 
\section{FAVis System Design}

As shown in Figure~\ref{fig:teaser}, FAVis has multiple views: matrix, network, parallel-coordinates views for variables and factors, as well as tag, word cloud, threshold, and factor correlation views. 
Scaling and dragging are supported for the objects in matrix, network, and factor correlation views.


{\bf Matrix view}
(Figure~\ref{fig:teaser}A) shows the factor loadings matrix $\bm{\Lambda}$ as a heatmap and as a matrix. Each rectangle represents a factor loading corresponding to its variable and factor (\textbf{G1}, \textbf{G2}). Each axis represents variables or factors depending on its \textit{transpose} setting. The color of rectangles is based on the values of factor loadings. Only the absolute values of factor loadings larger than the threshold value will be shown here. Displaying absolute or actual values is determined by its \textit{absolute values} settings. If users decide to show absolute values of factor loadings, only the degree of the associations between variables and factors will be displayed. On the other hand, if actual values are shown, positive values are displayed as light blue, and negative values are displayed as light red, signifying the direction of the correlation between its variable and factor. Variables and factors can be sorted by clicking a variable or factor of interest, which helps users identify the most relevant variables/factors to their corresponding factors/variables. The sorting will also be reflected on the parallel-coordinates view. 

{\bf Network view}
(Figure~\ref{fig:teaser}B) displays a graph of variables based on factor loadings (\textbf{G1}, \textbf{G2}) using the force-directed layout. Each node represents a variable, and an edge is formed if two variables have factor loadings in any factor larger than the threshold. This visualization is generated from the factor loadings matrix to highlight cross-loadings and eliminate redundant-loadings. Suppose $\lambda_{ik}$ is the factor loading for variable $i$ and factor $k$, and $\alpha$ is the threshold value. An edge between variable $i$ and variable $j$ is formed if $|\lambda_{ik}| > \alpha$ and $|\lambda_{jk}| > \alpha$ for any $k$. Multiple factors could satisfy this condition for variables $i$ and $j$. In such a case, these factor loadings are considered as redundant-loadings. Depending on a goal (\textbf{G1} or \textbf{G2}), users can choose between two modes of visualization in the network view. For $\textbf{G1}$, the color of nodes and edges are determined based on the largest factor loading. Specifically, the color of nodes will be determined by choosing the factor with the largest factor loading, and the color of edges will be determined using the factor with the largest average factor loadings of variables $i$ and $j$. Since the network view can further reduce the amount of information by eliminating redundant-loadings, it offers a more summarized visualization than the matrix view. For \textbf{G2}, both nodes and edges are colored based on the number of cross-loadings (which includes redundant-loadings). This mode can easily find an optimal threshold value as the number of cross-loadings is vividly visualized.  

{\bf Parallel-coordinates views} 
(Figure~\ref{fig:teaser}C) show the line plots of factor loadings. One displays variables on the $x$-axis and factors on the $y$-axis, and the other does the opposite. 
These views select variables/factors with the largest factor loadings on a specific factor/variable (\textbf{G4}). Users can create filters by dragging a mouse on a $y$-axis, highlighting a subset of variables/factors in all the other views. Since these views are not affected by the threshold, they are suitable for examining a specific variable or factor without information loss. To reduce visual clutter while maintaining task effectiveness, we only show one in every five $y$-axes when there are more than 20. The hidden ones will be shown when hovering over them.

\begin{figure}[htb]
\vspace{-0.1in}
\begin{center}
$\begin{array}{c@{\hspace{0.1in}}c}
		\includegraphics[width=0.45\linewidth]{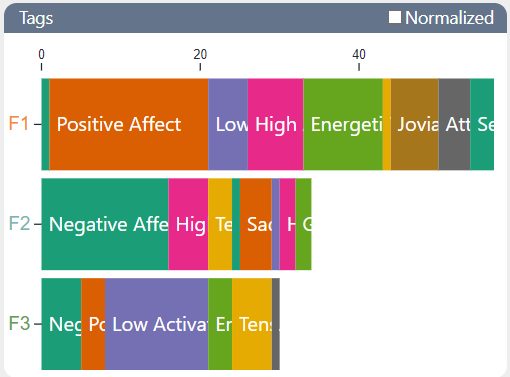} &
		\includegraphics[width=0.45\linewidth]{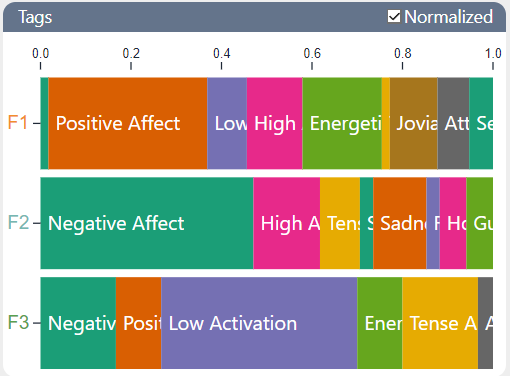} \\
		\mbox{\footnotesize (a) Count} & \mbox{\footnotesize (b) Proportion}
\end{array}$
\end{center}
   \vspace{-0.25in}
 \caption{Comparison of tag views: (a) showing counts and (b) showing proportions of variables with associated tags. (a) is beneficial for comparing the effectiveness of different factors in measuring theorized psychological constructs, while (b) aids in interpreting the meaning of individual factors in terms of these constructs.} 
 \label{fig:comptags}
\end{figure}

{\bf Tag view}
(Figure~\ref{fig:teaser}D) links a theory to the resultant model (\textbf{G5}) to enhance understanding of a CFM. Users can assign a tag representing a psychological construct of interest to a presumed related variable. They can add tags within the ``Edit Tag'' interface (Figure~\ref{fig:teaser}Da). This is done by entering the desired tag name into the input box and clicking the ``Add Tag'' button. To facilitate this process, users can automatically add tags using a codebook file (a dictionary specifying variables and their associated list of tags). Tags associated with a selected variable are highlighted by increasing their opacity. Users can also click on any tag to toggle its assignment to the selected variable. Then, the tag view displays these tags for each factor through a horizontal stacked bar chart (Figure~\ref{fig:teaser}Db). Each rectangle within this chart symbolizes a tag, with its width proportional to the count of associated variables exceeding the threshold. The design ensures only pertinent tags are shown, with each rectangle's size proportional to the tag's relevance to a factor, aiding in factor interpretation. When a rectangle's width is insufficient to display a tag's text, hovering over it fully activates a tooltip that reveals the tag's name and value (the rectangle's width). When the ``Normalized'' checkbox on the top bar of the tag view is checked, instead of the raw count, the width of a rectangle will become proportional to the total counts (Figure~\ref{fig:comptags}b), which is more helpful when users only want to interpret the meaning of a factor, instead of exploring potential values of the threshold. 

{\bf Word cloud view} 
(Figure~\ref{fig:teaser}E) is the most suitable for interpreting the meaning of a factor without relying on tags (\textbf{G1}). For a selected factor, each text element corresponds to the name of a variable, with its size and color determined by the factor loading. The color reflects the factor loading value, following the same color scale as the matrix view. Similarly, the font size is scaled linearly according to the absolute value of the factor loading. Therefore, users can interpret the meaning of a factor just by concentrating on a set of highlighted tags, which is particularly helpful when dealing with numerous variables, and the matrix view cannot display every variable simultaneously. 

{\bf Threshold view}
(Figure~\ref{fig:teaser}F) plots the cumulative distribution of the factor loadings and provides a slider to adjust the threshold used in the matrix and network views (\textbf{G3}). The threshold determines information loss to improve the interpretability of the model, as factor loadings below the threshold will not be shown in the matrix or network views. Information loss is depicted as the horizontal line in the plot, which corresponds to the portion of the factor loadings that will be discarded in the matrix and network views. The threshold value is shown as the vertical line. As users update the threshold value using the slider, the matrix and network views will also be dynamically updated. Users are expected to change the threshold value, which minimizes the number of cross-loadings, preserving most of the factor loadings  (\textbf{G3}). A good model is expected to have few cross-loadings even with a lower threshold value, as factors are well separated in such a case (\textbf{G2}).

{\bf Factor correlation view}
(Figure~\ref{fig:teaser}G) shows the correlations among factors using a graph with the force-directed layout (\textbf{G2}). Each node represents a factor, and each edge represents the correlation between two factors. This view will not display any edge if no factors are correlated, which can happen when we use PCA or orthogonal rotations.


{\bf Additional interactions and workflow.}
In the top bar (Figure~\ref{fig:teaser}H), there is a line input for searching variables by their names and filtering out these variables in the matrix, network, and parallel-coordinates views (\textbf{G4}). We can also limit the maximum number of variables or factors displayed in these views.
In FAVis, many interactions among views are implemented (\textbf{G4}). For example, suppose users click on a variable/factor on the matrix view. In that case, the factor loadings will be sorted by descending order, helping them identify the most relevant factors/variables, which will also be reflected in the parallel-coordinates views. Additionally, if users click a rectangle within the matrix view, the corresponding row and column are highlighted in red. This action also triggers highlights in the network and parallel-coordinates views, where irrelevant visual elements become less opaque, focusing attention on the selected variable and factor. Similarly, clicking a node in the network view highlights the selected variable consistently across other views. 
On hovering a rectangle in the matrix view or a node in the network view, a tooltip will appear, which contains relevant information about its corresponding variable and/or factor, such as its value of the factor loading, codebook information (e.g., question text), and associated tags to facilitate understanding of the results.

\vspace{-0.075in}
\section{Evaluation}

We conducted a user study to evaluate the usability of FAVis by following the procedures used in past studies \cite{bayrakPRAGMAInteractivelyConstructing2020,munechikaVisualAuditorInteractive2022}. Five domain experts (one professor, three quantitative psychology graduate students, and one statistics graduate student) familiar with EFA and/or PCA were selected to evaluate FAVis. The participants received no compensation for taking part in the study.

{\bf Data (CFMs).} Participants (denoted as $E1, E2, \ldots, E5$) were given two CFMs from two datasets during the study. The first CFM and dataset were a 12-factor CFM derived from a questionnaire for suicidal thought and behavior \cite{ken}, which were used for a tutorial on FAVis. The second CFM and dataset were a three-factor CFM derived from the Motivational State Questionnaire dataset \cite{revelle}, which consists of items taken from other scales~\cite{watsonDevelopmentValidationBrief1988, Thayer,larsenPromisesProblemsCircumplex1992}.

{\bf Procedure.} The study was conducted in person or via Zoom session (one participant). The participants were asked to (1) answer pre-study questions, (2) run FAVis in a browser and receive a tutorial using the first model, (3) reload the page to evaluate using the second model, and (4) answer post-study questions, including usability measures rated one to five. The participants were asked to conduct specific tasks related to the design goals. Since they did not have a theory for the dataset, they were first asked to evaluate the CFM without tags, and then we showed them pre-labeled results to re-evaluate the same CFM using tags. Each study session was recorded and lasted about 30 to 60 minutes.

{\bf Results.} Overall, all the participants rated the basic usability measures highly: an average of $4.1$ for \textit{Easy to Understand}, $4.4$ for \textit{Easy to Use}, $5.0$ for \textit{Enjoyable to Use}, and $5.0$ for \textit{Would use in the future}. The effectiveness of FAVis is also demonstrated by the task-specific measures: an average of $4.8$ for \textit{Helps you understand the meaning of each factor}, $5.0$ for \textit{Helps you understand factor relationships}, $4.4$ for \textit{Helps you identify the optimal threshold}, and $5.0$ for \textit{Helps you relate a theory to a model}. These results suggest that although FAVis is not necessarily the easiest to learn, as its average rating is the lowest, the overall satisfaction is very high. It is very useful to conduct these tasks. 

We mainly conducted three tasks: interpreting the meanings of factors with/without tags, explaining relationships among factors (e.g., cross-loadings and correlations), and identifying the best items for each factor. As demonstrated by the high average ratings of task-specific measures, all participants completed these tasks, accurately describing the meanings of factors and their relationships. However, there are differences among participants in approaching some of the tasks.

When identifying cross-loadings, \textit{E3}, \textit{E4}, and \textit{E5} kept adjusting the threshold until they found cross-loadings in the network view. In contrast, \textit{E1} used filters on the parallel-coordinates view to emphasize cross-loadings in other views. Both approaches are generally considered efficient for this task. Nevertheless, \textit{E2} took a different route, setting the threshold to zero to reveal all factor loadings on the matrix view, which was unexpected. This could be because some researchers are used to reading matrix values to the point where they find it challenging to unlearn their habitual methods, highlighting the possible difficulty in learning to use FAVis effectively. Regarding the parallel-coordinates views, \textit{E1}, who used the views for this task, said, ``It has a lot of potential but it is also the one that's got the highest ceiling for use.'' \textit{E1} also used the views for identifying the best items for each factor, saying, ``I think this tool is very powerful.'' On the other hand, \textit{E3} did not use them because, as he said, ``the views looked like spaghetti.''  This suggests that while the parallel-coordinates views have the range filtering feature, a more straightforward interface for range filtering might be preferable to some people. 
When interpreting the meaning of each factor, \textit{E5} especially preferred using the word cloud view (she mentioned that the word cloud view is instrumental when the names of variables are informative; otherwise, it is less useful), while others tended to use the matrix view. 

The usability measures show that all the participants enjoyed using the tool and liked different aspects, including the network view, thresholding, matrix view and sorting feature, and word cloud view. For example, \textit{E1} and \textit{E3} particularly liked the network view: \textit{E1} mentioned that he enjoyed watching the changes and movements of graphics when adjusting the threshold, and \textit{E3} said he enjoyed watching clusters and cross-loadings appearing in the network view as he changes the threshold.



\vspace{-0.075in}
\section{Conclusion, Limitations, and Future Work}

This study introduces FAVis, a visual analytics tool designed to help psychological researchers explore and understand CFMs. This research was motivated to fill a gap in the visualization literature, as few focus on visualizing behavioral data to identify psychological constructs. Having multiple views, FAVis can help users perform various tasks such as interpreting factors, understanding between-factor information, and relating a theory to a model. Our user study also revealed the tool's effectiveness in completing the different tasks, which led to high satisfaction. The multiple views allowed users with varying levels of familiarity with visual analytics and factor analysis to accomplish tasks using their preferred methods.


This study has several limitations. First, some users noted that the factor correlation view could benefit from supporting different visualization modalities. Currently, the view uses a force-directed graph, effective when factor correlations have more significant variability, leading to distinct graph shapes. However, the force-directed graph becomes less informative if the correlations exhibit less variability. In such a case, a heatmap representation would be more beneficial, as it provides a better interface to find a value associated with a specific pair of factors. Second, the tool's initial layout and ordering of the rows and columns in a matrix are fixed. FAVis should include a method to determine the optimal ordering for the rows and columns in a factor loadings matrix. Third, the tool cannot compare between different CFMs for the same dataset. Finally, the results from our user study primarily provide usability data and do not offer strong evidence of efficacy. Comparing this tool against a baseline method could be useful to demonstrate its efficacy. In the future, we aim to address these limitations to enhance FAVis.










\vspace{-0.075in}
\acknowledgments{This research was supported in part by the U.S.\ National Science Foundation through grants IIS-1955395, IIS-2101696, OAC-2104158, and IIS-2401144, and the U.S.\ Department of Energy through grant DE-SC0023145. The authors would like to thank the anonymous reviewers for their insightful comments.}

\vspace{-0.05in}
\bibliographystyle{abbrv-doi}

\bibliography{template}

\end{document}